\documentclass[twocolappendix,appendixfloats,]{emulateapj}

\usepackage[colorlinks = true, linkcolor = blue, urlcolor  = blue, citecolor = blue, anchorcolor = blue]{hyperref}
\usepackage{apjfonts}
\usepackage{rotating}
\usepackage{amsmath}
\usepackage{color}
\usepackage{graphicx}
\usepackage[dvipsnames]{xcolor}
\usepackage{CJK}

\newcommand{\pivec}{\mbox{\boldmath $\pi$}}
\newcommand{\muvec}{\mbox{\boldmath $\mu$}}

\newcommand{\thetae}{\theta_{\rm E}}
\newcommand{\pie}{\pi_{\rm E}}
\newcommand{\pien}{\pi_{{\rm E},N}}
\newcommand{\piee}{\pi_{{\rm E},E}}

\definecolor{darkbrown}{RGB}{139,69,19}

\shorttitle{Microlensing Planet in Binary}
\shortauthors{Han et al.}

\begin{document}

\title{OGLE-2018-BLG-1700L: Microlensing Planet in Binary Stellar System}

\author{
Cheongho~Han\altaffilmark{0001}, 
Chung-Uk~Lee\altaffilmark{0002,101}, 
Andrzej~Udalski\altaffilmark{0003,100}, 
Andrew~Gould\altaffilmark{0004,0005,101}, 
Ian~A.~Bond\altaffilmark{0006,102}
\\
(Leading authors),\\
and \\
Michael~D.~Albrow\altaffilmark{0007}, Sun-Ju~Chung\altaffilmark{0002,0008},  
Kyu-Ha~Hwang\altaffilmark{0002}, Youn~Kil~Jung\altaffilmark{0002}, 
Yoon-Hyun~Ryu\altaffilmark{0002}, In-Gu~Shin\altaffilmark{0002}, 
Yossi~Shvartzvald\altaffilmark{0009}, 
Jennifer~C.~Yee\altaffilmark{0010}, Weicheng Zang\altaffilmark{0011},
Sang-Mok~Cha\altaffilmark{0002,0012}, Dong-Jin~Kim\altaffilmark{0002}, Hyoun-Woo~Kim\altaffilmark{0002}, 
Seung-Lee~Kim\altaffilmark{0002,0008}, Dong-Joo~Lee\altaffilmark{0002}, Yongseok~Lee\altaffilmark{0002,0012}, 
Byeong-Gon~Park\altaffilmark{0002,0008}, Richard~W.~Pogge\altaffilmark{0005}, 
M.~James~Jee\altaffilmark{0013,0014}, Doeon~Kim\altaffilmark{0001},
\\
(The KMTNet Collaboration),\\
Przemek~Mr{\'o}z\altaffilmark{0003}, Micha{\l}~K.~Szyma{\'n}ski\altaffilmark{0003}, Jan~Skowron\altaffilmark{0003},
Radek~Poleski\altaffilmark{0005}, Igor~Soszy{\'n}ski\altaffilmark{0003}, Pawe{\l}~Pietrukowicz\altaffilmark{0003},
Szymon~Koz{\l}owski\altaffilmark{0003}, Krzysztof~Ulaczyk\altaffilmark{0015}, Krzysztof~A.~Rybicki\altaffilmark{0003},
Patryk~Iwanek\altaffilmark{0003}, Marcin~Wrona\altaffilmark{0003}\\
(The OGLE Collaboration) \\   
Fumio~Abe\altaffilmark{0016}, Richard~Barry\altaffilmark{0017},           
David~P.~Bennett\altaffilmark{0017,0018},
Aparna~Bhattacharya\altaffilmark{0017,0018}, Martin~Donachie\altaffilmark{0019}, 
Hirosane~Fujii\altaffilmark{0016},
Akihiko~Fukui\altaffilmark{0020,0021},  Yoshitaka~Itow\altaffilmark{0016}, 
Yuki~Hirao\altaffilmark{0022}, 
Yuhei~Kamei\altaffilmark{0016},
Iona~Kondo\altaffilmark{0022}, 
Naoki~Koshimoto\altaffilmark{0023,0024}, Man~Cheung~Alex~Li\altaffilmark{0019},    
Yutaka~Matsubara\altaffilmark{0016}, Yasushi~Muraki\altaffilmark{0016}, 
Shota~Miyazaki\altaffilmark{0022}, Masayuki~Nagakane\altaffilmark{0022}, 
Cl\'ement~Ranc\altaffilmark{0017}, Nicholas~J.~Rattenbury\altaffilmark{0019}, 
Haruno~Suematsu\altaffilmark{0022}, Denis~J.~Sullivan\altaffilmark{0025}, 
Takahiro~Sumi\altaffilmark{0022},  Daisuke~Suzuki\altaffilmark{0026}, 
Paul~J.~Tristram\altaffilmark{0027}, 
Takeharu~Yamakawa\altaffilmark{0016},
Atsunori~Yonehara\altaffilmark{0028}\\         
(The MOA Collaboration),\\ 
}

\email{cheongho@astroph.chungbuk.ac.kr}

\altaffiltext{0001}{Department of Physics, Chungbuk National University, Cheongju 28644, Republic of Korea} 
\altaffiltext{0002}{Korea Astronomy and Space Science Institute, Daejon 34055, Republic of Korea} 
\altaffiltext{0003}{Warsaw University Observatory, Al.~Ujazdowskie 4, 00-478 Warszawa, Poland} 
\altaffiltext{0004}{Max Planck Institute for Astronomy, K\"onigstuhl 17, D-69117 Heidelberg, Germany} 
\altaffiltext{0005}{Department of Astronomy, Ohio State University, 140 W. 18th Ave., Columbus, OH 43210, USA} 
\altaffiltext{0006}{Institute of Natural and Mathematical Sciences, Massey University, Auckland 0745, New Zealand}
\altaffiltext{0007}{University of Canterbury, Department of Physics and Astronomy, Private Bag 4800, Christchurch 8020, New Zealand} 
\altaffiltext{0008}{Korea University of Science and Technology, 217 Gajeong-ro, Yuseong-gu, Daejeon, 34113, Republic of Korea} 
\altaffiltext{0009}{IPAC, Mail Code 100-22, Caltech, 1200 E.\ California Blvd., Pasadena, CA 91125, USA}
\altaffiltext{0010}{Center for Astrophysics $\|$ Harvard \& Smithsonian 60 Garden St., Cambridge, MA 02138, USA} 
\altaffiltext{0011}{Physics Department and Tsinghua Centre for Astrophysics, Tsinghua University, Beijing 100084, China} 
\altaffiltext{0012}{School of Space Research, Kyung Hee University, Yongin, Kyeonggi 17104, Korea} 
\altaffiltext{0013}{Yonsei University, Department of Astronomy, Seoul, Republic of Korea}
\altaffiltext{0014}{Department of Physics, University of California, Davis, California, USA}
\altaffiltext{0015}{Department of Physics, University of Warwick, Gibbet Hill Road, Coventry, CV4 7AL, UK} 
\altaffiltext{0016}{Institute for Space-Earth Environmental Research, Nagoya University, Nagoya 464-8601, Japan}
\altaffiltext{0017}{Code 667, NASA Goddard Space Flight Center, Greenbelt, MD 20771, USA}
\altaffiltext{0018}{Department of Astronomy, University of Maryland, College Park, MD 20742, USA}
\altaffiltext{0019}{Department of Physics, University of Auckland, Private Bag 92019, Auckland, New Zealand}
\altaffiltext{0020}{Instituto de Astrof\'isica de Canarias, V\'ia L\'actea s/n, E-38205 La Laguna, Tenerife, Spain}
\altaffiltext{0021}{Department of Earth and Planetary Science, Graduate School of Science, The University of Tokyo, 7-3-1 Hongo, Bunkyo-ku, Tokyo 113-0033, Japan}
\altaffiltext{0022}{Department of Earth and Space Science, Graduate School of Science, Osaka University, Toyonaka, Osaka 560-0043, Japan}
\altaffiltext{0023}{Department of Astronomy, Graduate School of Science, The University of Tokyo, 7-3-1 Hongo, Bunkyo-ku, Tokyo 113-0033, Japan}
\altaffiltext{0024}{National Astronomical Observatory of Japan, 2-21-1 Osawa, Mitaka, Tokyo 181-8588, Japan}
\altaffiltext{0025}{School of Chemical and Physical Sciences, Victoria University, Wellington, New Zealand}
\altaffiltext{0026}{Institute of Space and Astronautical Science, Japan Aerospace Exploration Agency, 3-1-1 Yoshinodai, Chuo, Sagamihara, Kanagawa, 252-5210, Japan}
\altaffiltext{0027}{University of Canterbury Mt.\ John Observatory, P.O. Box 56, Lake Tekapo 8770, New Zealand}
\altaffiltext{0028}{Department of Physics, Faculty of Science, Kyoto Sangyo University, 603-8555 Kyoto, Japan}
\altaffiltext{100}{OGLE Collaboration.}
\altaffiltext{101}{KMTNet Collaboration.}
\altaffiltext{102}{MOA Collaboration.}


\begin{abstract}
We report the discovery of a planet in a binary that was discovered from the analysis 
of the microlensing event OGLE-2018-BLG-1700.  We identify the triple nature of the lens 
from the fact that the complex anomaly pattern can be decomposed into two parts produced 
by two binary-lens events, in which one binary pair has a very low mass ratio of $\sim 0.01$ 
between the lens components and the other pair has a mass ratio of $\sim 0.3$.  We find two 
sets of degenerate solutions, in which one solution has a projected separation between the 
primary and its stellar companion less than the angular Einstein radius $\thetae$ (close 
solution), while the other solution has a separation greater than $\thetae$ (wide solution).  
From the Bayesian analysis with the constraints of the event time scale and angular Einstein 
radius together with the location of the source lying in the far disk behind the bulge, we 
find that the planet is a super-Jupiter with a mass of $4.4^{+3.0}_{-2.0}~M_{\rm J}$ and 
the stellar binary components are early and late M-type dwarfs with masses 
$0.42^{+0.29}_{-0.19}~M_\odot$ and $0.12^{+0.08}_{-0.05}~M_\odot$, respectively, and the 
planetary system is located at a distance of $D_{\rm L}=7.6^{+1.2}_{-0.9}~{\rm kpc}$.  The 
planet is a circumstellar planet according to the wide solution, while it is a circumbinary 
planet according to the close solution.  The projected primary-planet separation is 
$2.8^{+3.2}_{-2.5}~{\rm au}$ commonly for the close and wide solutions, but the primary-secondary 
binary separation of the close solution, $0.75^{+0.87}_{-0.66}~{\rm au}$, is widely different 
from the separation, $10.5^{+12.1}_{-9.2}~{\rm au}$, of the wide solution.
\end{abstract}


\section{Introduction}\label{sec:one}

Since the first-generation microlensing experiments conducted in the early 1990s, 
e.g., MACHO \citep{Alcock1993}, EROS \citep{Aubourg1993}, and OGLE-I \citep{Udalski1994}, 
the detection rate of microlensing events has dramatically increased.  Compared to the 
rate of several dozens per year in the early stage, current lensing experiments, OGLE-IV 
\citep{Udalski2015}, MOA \citep{Bond2001}, and KMTNet \citep{Kim2016}, annually report 
more than 3000 events.  The greatly enhanced detection rate has become possible thanks 
to the increased monitoring cadence with the use of multiple telescopes equipped with 
large-format cameras.

With the increase of the event rate, the number of anomalous events, which exhibit
deviations in lensing light curves from the standard form of a single-lens (1L)
single-source (1S) event, has also increased. The most common case of anomalous
events is binary-lens events, in which a single source is gravitationally lensed by a
binary lens composed of two masses (2L1S). Binary-lens events are produced by various
combinations of astronomical objects. As expected from the high stellar binary rate,
the majority of 2L1S events are produced by binaries that are composed of two stars 
with similar masses.  Binary-lens events are also produced by the star-planet combination, 
and this makes microlensing an important tool to detect extrasolar planets 
\citep{Mao1991, Gould1992b}, especially those located around and beyond the snow lines 
of faint M dwarfs.

Although not very common, the number of events produced by triple lenses (3L1S
events) is also increasing. By the time of writing this paper, there are nine
published 3L1S events. Among them, five events were produced by multiplanet systems,
including
OGLE-2006-BLG-109 \citep{Gaudi2008, Bennett2010},
OGLE-2012-BLG-0026 \citep{Han2013, Beaulieu2016},
OGLE-2014-BLG-1722 \citep{Suzuki2018},
OGLE-2018-BLG-0532 \citep{Ryu2019}, and
OGLE-2018-BLG-1011 \citep{Han2019a}.\footnote{We note that the signals of 
two planets are securely detected for events OGLE-2006-BLG-109, OGLE-2012-BLG-0026, 
and OGLE-2018-BLG-1011.  However, the signals of the second planets for the events 
OGLE-2014-BLG-1722 and OGLE-2018-BLG-0532 are rather less secure.} We note that all 
of these microlensing multiplanetary systems were detected through the channel of 
central perturbations, in which the source passes close to the central magnification 
region around the host star of the planets \citep{Griest1998}.  The high detection 
efficiency of this channel originates in the properties of lensing caustics induced 
by planetary companions.  A planetary companion located around the Einstein ring of 
the host induces two sets of caustics, in which one set is located close to the host 
(central caustic) and the other set is positioned away from the host (planetary caustic). 
If a lens contains multiple planets, the individual planets induce central caustics in 
the common central region and affect the magnification pattern of the region. Then, the 
chance to detect multiple planets is high for high-magnification events produced by the 
source approach close to the host of the planet \citep{Gaudi1998}.

Another population of the known triple-lens events are those produced by planets 
in binaries. These events include
OGLE-2007-BLG-349 \citep{Bennett2016},
OGLE-2008-BLG-092 \citep{Poleski2014},
OGLE-2013-BLG-0341 \citep{Gould2014}, and
OGLE-2016-BLG-0613 \citep{Han2017}.
For OGLE-2008-BLG-092 and OGLE-2013-BLG-0341, the planets were identified by their own 
independent signals.  Besides this independent channel, planets in binary systems 
can also be found through the central perturbation channel.  This is possible because 
both planet and binary companion can induce caustics in a common region, which is the 
region around the planet-hosting binary star for a S-type planet (circumstellar planet) 
orbiting around one of the two widely separated binary stars and the region around the 
barycenter of the binary for a P-type planet (circumbinary planet) orbiting around the 
center of mass of the closely located binary stars. The microlensing planets 
OGLE-2007-BLG-349L(AB)c and OGLE-2016-BLG-0613L(AB)c were detected though this central 
perturbation channel.

Besides multiple-planetary systems and planetary systems in binaries, triple lensing can
also provide channels to probe various types of astronomical systems, such as triple stars
and stars with a planet and a moon \citep{Han2002, Han2008, Liebig2010}.  From the analysis 
of the lensing event OGLE-2015-BLG-1459, \citet{Hwang2018} pointed out the possibility that 
the lens of the event was composed of a brown dwarf host, a Neptune-class planet, and a 
third body being a Mars-class object that could have been a moon of the planet.

Despite the usefulness in studying various astronomical objects, application of triple 
lensing is often hindered by the difficulty of analyzing events.  This difficulty 
arises because triple-lens systems exhibit very complex caustic patterns such as nested 
and self-intersected caustics, and this results in lensing light curves of great diversity.  
Theoretically, the ranges of the critical curve topology and the caustic structure have 
not yet been fully explored, and thus the understanding about the lensing behavior of 
triple-lens systems is still incomplete \citep{Rhie2002, Danek2015, Danek2019}.

Fortunately, triple-lensing events can be readily analyzed for events produced by 
some specific cases of lens systems.  These are the cases in which the 3L1S anomaly 
in the lensing light curve can be approximated by the superposition of the anomalies 
produced by two 2L1S events.  \citet{Bozza1999} and \citet{Han2001} pointed out that this 
superposition approximation could be used to analyze central perturbations induced 
by multiple planets. \citet{Lee2008} indicated that the approximation could also be 
applied for the detections and characterizations of planets in binary systems.

In this paper, we report the discovery of a new planet that belongs to a stellar 
binary system.  The planetary system was found from the analysis of the microlensing 
event OGLE-2018-BLG-1700.  The light curve of the event exhibits a complex pattern 
with multiple anomaly features.  We identify the triple nature of the lens from the 
fact that the anomaly pattern can be decomposed into two parts produced by two 2L1S 
events.

The paper is organized as follows.  In Section~\ref{sec:two}, we mention the acquisition 
and processing of data used in the analysis.  In Section~\ref{sec:three}, we describe the 
analysis process that leads to the identification of the planet in a binary.  We also 
present local solutions resulting from degeneracies.  In Section~\ref{sec:four}, we 
characterize the source from its color and brightness.  In Section~\ref{sec:five}, we 
estimate the physical lens parameters including the mass and distance to the lens.  
We summarize results and conclude in Section~\ref{sec:six}.

\section{Observation and Data}\label{sec:two}

The source star of the lensing event OGLE-2018-BLG-1700 is located toward the Galactic 
bulge field with the equatorial coordinates $({\rm RA}, {\rm decl.})= (17:59:49.45, 
-28:31:43.1)$. The corresponding Galactic coordinates of the source are 
$(l,b)=(1^\circ\hskip-2pt.93, -2^\circ\hskip-2pt.47)$.  The apparent baseline magnitude 
of the source is $I_{\rm base}=17.65$, but as we will show in Section~\ref{sec:four}, 
the source is heavily blended and it comprises only $\sim 9\%$ of the baseline flux.

The lensing event was 
first found by the Optical Gravitational Microlensing Experiment \citep[OGLE:][]{Udalski2015} 
survey on 2018-09-15 (${\rm HJD}^\prime\equiv {\rm HJD}-2450000\sim 8376$), which corresponded 
to the early stage of the source-flux brightening. The OGLE survey was conducted using the 
1.3 m Warsaw Telescope located at the Las Campanas Observatory in Chile.  OGLE observations 
of the event were done with a cadence of $\sim 2$--3/night using $I$- and $V$-band filters.

\begin{figure}
\includegraphics[width=\columnwidth]{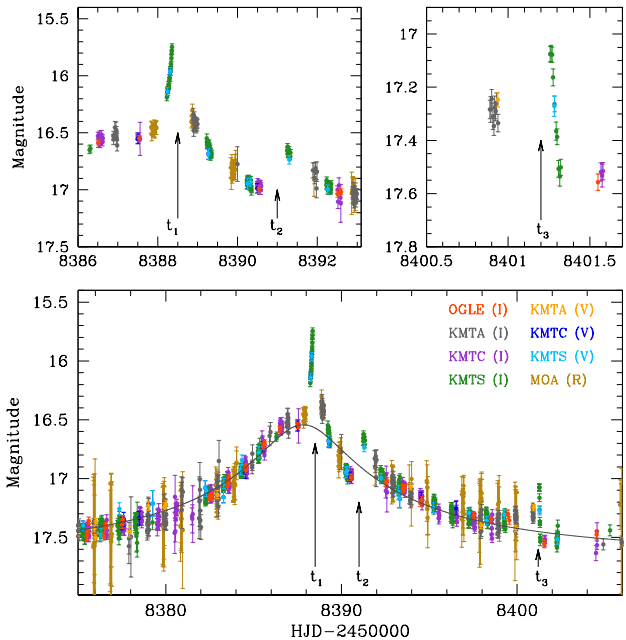}
\caption{
Light curve of the microlensing event OGLE-2018-BLG-1700.  The lower panel shows the 
whole view of the event and the upper panels show zooms of the regions around the 
three peaks at the times marked by $t_1$, $t_2$, and $t_3$.  The colors of the labels 
for the telescopes used for observations match those of the data points.  The curve 
superposed on the data points is the model obtained from single-lens and single-source 
(1L1S) fitting of the data excluding the data points around the anomaly peak at $t_1$.  
\bigskip
}
\label{fig:one}
\end{figure}

The event was also observed by the Korea Microlensing Telescope Network 
\citep[KMTNet:][]{Kim2016} survey.  The KMTNet survey was conducted utilizing three 
identical 1.6~m telescopes at the Siding Spring Observatory, Australia, Cerro Tololo 
Interamerican Observatory, Chile, and the South African Astronomical Observatory, South 
Africa. Hereafter, we refer to the individual KMTNet telescopes as KMTA, KMTC, and KMTS, 
respectively. The event was independently found from the analysis of the 2018 data 
conducted after the season \citep{Kim2018} and it was designated as KMT-2018-BLG-2330.  
KMTNet observations of the event were carried out mostly in $I$ band with a 
15-min cadence for each telescope.  Some $V$-band data were obtained mainly for the 
purpose of measuring the source color, but in our analysis, we include them in the 
analysis to maximize the coverage of the light curve.  The KMTNet $V$-band data were 
obtained with a cadence corresponding to $\sim 1/10$ of the $I$-band cadence.

There exist additional data from the Microlensing Observations in Astrophysics 
\citep[MOA:][]{Bond2001, Sumi2003} survey.  The event was not alerted by the MOA survey 
but it was located in the middle of the their high-cadence fields. The MOA data were 
produced from the post-season photometry conducted for the source star found by other 
surveys.  The MOA survey was done in a customized broad $R$ band utilizing the 1.8~m 
telescope of the Mt.~John Observatory in New Zealand.

Data used in the analysis are processed using the photometry codes developed based 
on the difference imaging technique \citep{Alard1998} and customized by the individual 
survey groups: \citet{Wozniak2000} (OGLE), \citet{Albrow2017} (KMTNet), and \citet{Bond2001} 
(MOA).  We normalize the error bars of the individual data sets using the method of 
\citet{Yee2012}. For a subset of KMTNet data (KMTC), we conduct additional photometry 
using the pyDIA photometry code \citep{Albrow2017} to measure the source color.

In Figure~\ref{fig:one}, we present the light curve of the event constructed with the 
combined data. The curve superposed on the data points in the lower 
panel shows the 1L1S model obtained by fitting the 
data excluding the data points around the anomaly peak at ${\rm HJD}^\prime\sim 8388$.  
The light curve shows a complex pattern of deviation from the 1L1S model. The deviation 
is characterized by three peaks that are centered at ${\rm HJD}^\prime\sim 8388.2$ ($t_1$), 
8390.9 ($t_2$), and 8401.2 ($t_3$). We mark the individual peaks with arrows. In the 
upper two panels, we present the enlarged views of the peaks.  The peaks at $t_2$ and 
$t_3$ together with the U-shape trough region between the peaks indicate that these 
peaks are produced by caustic crossings, in which the former and latter peaks occur 
when the source enters and exits the closed curve of a binary caustic, respectively. 
The peak at $t_1$, on the other hand, does not show a counterpart peak of the 
caustic-crossing pair. This suggests that the peak is likely to be produced by the source 
approach close to the cusp of a caustic.

\begin{figure}
\includegraphics[width=\columnwidth]{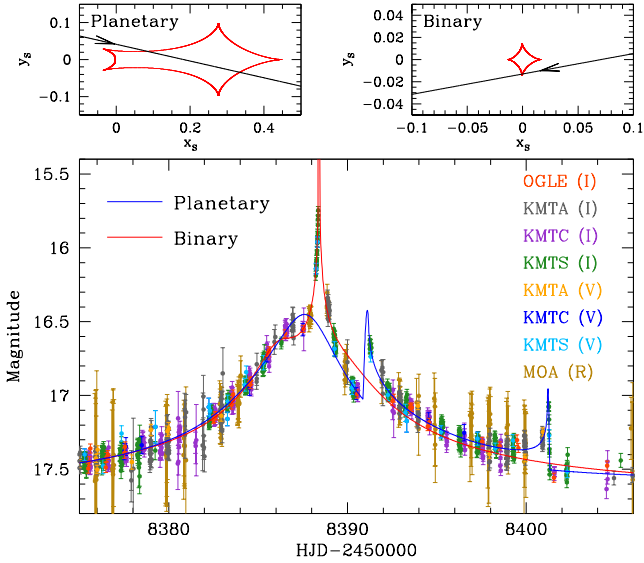}
\caption{
Decomposition of the anomaly into two parts produced by two binary-lens single-source 
(2L1S) events.  The blue and red curves are the models of the 2L1S solutions obtained 
by fitting two sets of data, for each of which a part of the data is excluded.  For 
the first data set, the data in the region 
$8387.0 < {\rm HJD}^\prime\equiv {\rm HJD}-2450000 < 8395.5$ are excluded, while for 
the second set, the data in the region $8389.5 < {\rm HJD}^\prime < 8405.0$ are excluded.  
The 2L1S fit to the first data set results in a 2L1S solution (blue curve) with a very 
low mass ratio of $q\sim 0.01$, and thus we designate the model as ``planetary''.  The 
fit to the second data set (red curve), on the other hand, results in $q\sim 0.3$, and 
thus the solution is designated as ``binary''.  The upper panels show the lens-system 
configurations of the planetary (left panel) and binary (right panel) solutions.  For 
each panel, the closed concave curve represents the caustic and the line with an arrows 
indicates the source trajectory.
\bigskip
}
\label{fig:two}
\end{figure}

\section{Light Curve Modeling}\label{sec:three}

\subsection{2L1S Analysis}\label{sec:three-one}

Because the anomaly features in the light curve are likely to be involved with 
caustics, we start the modeling of the observed light curve with a model, in which a 
single source is lensed by a binary lens (2L1S).  In 2L1S modeling, a basic 
description of the lensing light curve requires 7 lensing parameters, including 
$t_0$, $u_0$, $t_{\rm E}$, $s$, $q$, $\alpha$, and $\rho$. The first three parameters 
($t_0$, $u_0$, $t_{\rm E}$) represent the time of the closest approach of the source 
to a reference position of the lens, the source-reference separation at that time, 
and the event timescale, respectively. We use the center of mass as a reference position 
of the lens. The parameters $(s, q)$ denote the projected binary separation and the 
companion/primary mass ratio, respectively, and $\alpha$ represents the angle between 
the source trajectory and the binary lens axis. We note that the lengths of $u_0$ and 
$s$ are normalized to the angular Einstein radius $\thetae$. The last parameter $\rho$ 
indicates the ratio of the angular source radius $\theta_*$ to $\thetae$, i.e., 
$\rho=\theta_*/\thetae$ (normalized source radius). The normalized source radius is 
needed to describe the caustic-crossing parts, during which the lensing magnifications 
are affected by finite-source effects.

\begin{figure*}
\epsscale{0.93}
\plotone{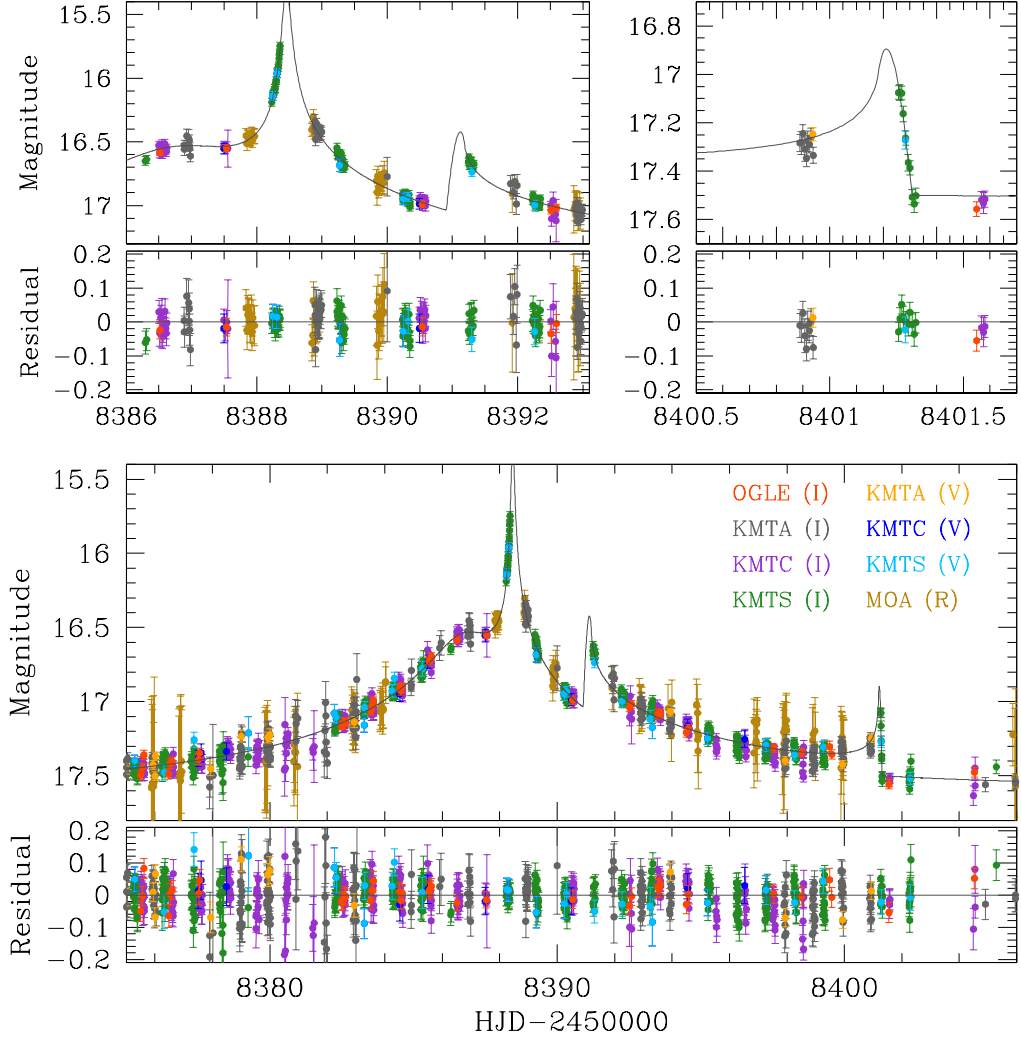}
\caption{
Model curve of 3L1S solution. The upper panels show the enlarged views of the peak 
regions.  The presented model is for the wide solution, in which the separation between 
$M_1$ and $M_3$ is greater than the Einstein radius, i.e., $s_3 > 1.0$. We note that the 
model curve of the ``close'' solution with $s_3 < 1.0$ is almost identical to the presented 
model curve of the ``wide'' solution.  
\bigskip
}
\label{fig:three}
\end{figure*}

Binary-lens modeling is conducted in two steps. In the first step, we conduct a grid 
search for the parameters $s$ and $q$, while the other parameters are searched for 
using a downhill approach based on the Markov Chain Monte Carlo (MCMC) algorithm. Once 
a plausible local solution is found from this first-round search, we then refine the 
solution by allowing all lensing parameters to vary.

We find that 2L1S modeling does not yield a model explaining all the anomaly features
despite repeated modeling runs with various combinations of the initial lensing parameters.
In order to check the possibility that the anomaly could be described with higher-order effects,
we consider two higher-order effects, including the microlens-lens parallax and the 
lens-orbital effects.  The former effects occur due to the orbital motion of Earth (observer) 
around the Sun \citep{Gould1992a} and the latter effects arise due to the orbital motion 
of the binary lens \citep{Dominik1998}.  Consideration of the microlens-parallax effect 
requires to include two additional lensing parameters of $\pien$ and $\piee$, which are 
the north and east components of the projected microlens-parallax vector, $\pivec_{\rm E}$, 
in the equatorial coordinates, respectively.  Consideration of the lens-orbital effects also 
requires to include two additional parameters of $ds/dt$ and $d\alpha/dt$, which denote the 
instantaneous change rates (at $t_1$) of the binary separation and source trajectory angle, 
respectively.  From these additional modeling runs, it is found that the anomaly features 
cannot be explained even with these higher-order effects.

\subsection{3L1S Analysis}\label{sec:three-two}

Not being able to explain the light curve with 2L1S models, we then consider models, in 
which the lens is composed of three masses (3L).  With the introduction of a third body 
$M_3$ in addition to the binary lens components of $M_1$ and $M_2$, one needs to include 
additional lensing parameters. These parameters are the separation of the third body from 
the primary $M_1$, $s_3$, the mass ratio $q_3=M_3/M_1$, and the orientation angle of $M_3$ 
as measured from the $M_1$--$M_2$ axis, $\psi$.  We use the notations $s_2$ and $q_2$ to 
denote the $M_1$--$M_2$ separation and $M_2/M_1$ mass ratio, respectively.

Due to the large number of the 3L1S lensing parameters, which reaches 10, i.e.,  
($t_0, u_0, t_{\rm E}, s_2, q_2, \alpha, s_3, q_3, \psi, \rho)$, not even considering 
higher-order effects, it is difficult to explore all of the parameter space.  We, therefore, 
check the possibility of using the ``binary superposition'' approximation, in which the 
anomalies in the triple-lensing light curve is approximated by the superposition of the 
anomalies produced by the two hypothetical binary-lensing events that would be produced 
by the $M_1$--$M_2$ and $M_1$--$M_3$ pairs.  Under this approximation, we conduct 2L1S 
modeling for two sets of data, for each of which a part of the data is excluded.  In the 
first data set, we exclude the data in the region $8387.0 < {\rm HJD}^\prime < 8389.5$, 
which corresponds to the region around the first anomaly centered at $t_1$.  In the second 
data set, we exclude the data in the region $8389.5 < {\rm HJD}^\prime < 8405.0$, within 
which the pair of the caustic crossing peaks at $t_2$ and $t_3$ are included.

In Figure~\ref{fig:two}, we present the two model light curves obtained from 2L1S fitting 
to the two separate data sets.  The blue curve represents the model obtained from 2L1S 
fitting to the data set excluding the region around the peak at $t_1$, and the red 
curve is the model obtained from fitting to the data set excluding the caustic-crossing 
spikes at $t_2$ and $t_3$.  We find that the anomalies are decomposed into two parts 
produced by the two 2L1S events, in which the blue model curve well describes the anomalies 
in the region including $t_2$ and $t_3$, while the red model curve explains the peak at 
$t_1$.  This indicates that the event is produced by a lens with triple components and the 
anomaly in the lensing light curve can be well described by the ``binary superposition'' 
approximation.  The binary parameters corresponding to the blue model curve are
$(s,q)\sim (1.1, 0.01)$, indicating that the companion $M_2$ is a planetary mass object 
located near the Einstein radius of the primary lens component $M_1$.  For the model 
of the red curve, on the other hand, the mass ratio of the companion to the primary is 
$q\sim 0.3$, indicating that the third body $M_3$ is a stellar companion to the primary.  
We refer to the individual binary solutions as ``planetary'' and ``binary'' solutions, 
respectively.  For the $M_1$--$M_3$ binary pair, we find two solutions, in which one 
solution has a separation between the binary components much smaller than the Einstein 
radius ($s\ll 1.0$) and the other solution has a separation much greater than the Einstein 
radius ($s\gg 1.0$).

\begin{deluxetable}{lcc}
\tablecaption{Best-fit lensing parameters\label{table:one}}
\tablewidth{240pt}
\tablehead{
\multicolumn{1}{c}{Parameter}            &
\multicolumn{1}{c}{Wide ($s_3>1.0$)}     &
\multicolumn{1}{c}{Close ($s_3<1.0$)}
}
\startdata                                              
$t_0$ (${\rm HJD}^\prime$) &  8386.152 $\pm$ 0.040   &  8385.827 $\pm$ 0.065 \\
$u_0$ ($10^{-3}$)          &  5.88 $\pm$ 0.66        &  6.70 $\pm$ 0.87      \\
$t_{\rm E}$ (days)         &  43.12 $\pm$ 0.74       &  41.91 $\pm$ 0.82     \\
$s_2$                      &  1.019 $\pm$ 0.003      &  1.184 $\pm$ 0.003    \\
$q_2$                      &  0.010 $\pm$ 0.001      &  0.010 $\pm$ 0.001    \\
$\alpha$ (rad)             &  3.432 $\pm$ 0.007      &  3.368 $\pm$ 0.007    \\
$s_3$                      &  3.823 $\pm$ 0.022      &  0.274 $\pm$ 0.003    \\
$q_3$                      &  0.274 $\pm$ 0.010      &  0.297 $\pm$ 0.009    \\
$\psi$ (rad)               &  5.525 $\pm$ 0.014      &  5.625 $\pm$ 0.015    \\
$\rho$ ($10^{-3}$)         &  1.00 $\pm$ 0.07        &  0.95 $\pm$ 0.07 
\enddata                            
\tablecomments{${\rm HJD}^\prime={\rm HJD-2450000}$.
}
\end{deluxetable}

\begin{figure}
\includegraphics[width=\columnwidth]{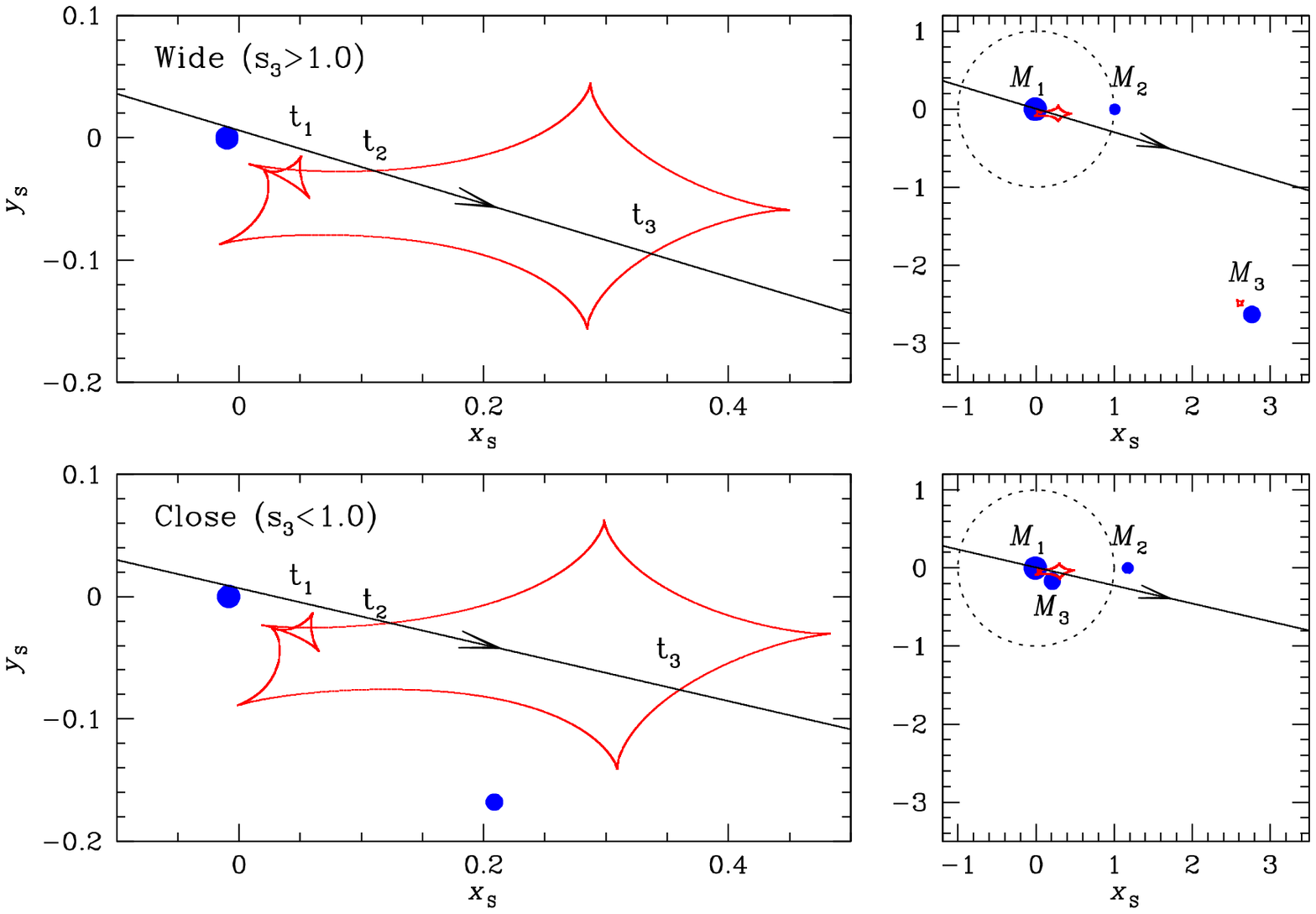}
\caption{
Lens-system configurations of the 3L1S solutions.  The upper and lower panels are for 
the wide ($s_3>1.0$) and close ($s_3<1.0$) solutions, respectively. For each solution, 
the left panel shows the central magnification region, while the right panel shows the 
whole view including the locations of all lens components, marked by $M_1$, $M_2$, and 
$M_3$.  The positions on the source trajectory marked by $t_1$, $t_2$, and $t_3$ represent 
the source locations at the times of the three peaks in the lensing light curve marked 
in Fig.~\ref{fig:one}.  The dotted circle in each of the right panels represents the 
Einstein ring.
\bigskip
}
\label{fig:four}
\end{figure}

In the two upper panels of Figure~\ref{fig:two}, we present the lens system configurations 
of the ``planetary'' and ``binary'' 2L1S solutions. In each panel, the closed figure 
composed of concave curves represents the caustic and the line with an arrow represents 
the source trajectory.  The caustics of the planetary solution form a single resonant 
hexalateral curve produced by a planetary companion.  On the other hand, the caustics of 
the binary solution form a concave quadrilateral curve.

Using the lensing parameters of the two 2L1S solutions as initial parameters, we then 
conduct 3L1S modeling. In Figure~\ref{fig:three}, we present the best-fit 3L1S model 
curve superposed on the observed data points. It is found that the 3L1S solution well 
describes all the anomaly features.  In Table~\ref{table:one}, we present the lensing 
parameters of the 3L1S solution.  We find that there exist two solutions resulting from 
the close/wide degeneracy in the $M_1$--$M_3$ separation, i.e., $s_3$, but we note that 
there is no close/wide degeneracy in the $M_1$--$M_2$ separation, i.e., $s_2$, because 
$s_2\sim 1.0$ and thus the $M_1$--$M_2$ binary pair forms a resonant caustic.  We note 
that the corresponding lensing parameters of the pair of degenerate solutions are similar 
to each other except that $s_{3,{\rm close}}\sim 1/s_{3,{\rm wide}}$. Hereafter, we 
designate the solutions with $s_3>1.0$ and $s_3<1.0$ as ``wide'' and ``close'' solutions, 
respectively.  The degeneracy between the two solutions is relatively severe, with 
$\Delta\chi^2=2.7$.

In Figure~\ref{fig:four}, we present the lens-system configurations of the 3L1S solutions, 
in which the upper and lower panels are for the wide and close solutions, respectively. 
For each of the solutions, the left panel shows the central magnification region, while 
the right panel shows the whole view including the locations of all lens components.  
As expected from the severe degeneracy between the wide and close solutions, the lens-system 
configurations in the central region of the two solutions are very similar to each other.  
From the investigation of the configurations, it is found that the overall pattern of the 
central caustic is similar to the resonant caustic produced by the $M_1$--$M_2$ pair of 
the 2L1S planetary solution, presented in the upper left panel of Figure~\ref{fig:two}.  
The source passes the caustic diagonally, crossing the upper left and lower right folds 
of the caustic, thereby producing the peaks at $t_2$ and $t_3$.  The difference of the 
triple-lens caustic from that of the planetary 2L1S solution is that there exists a 
triangular-shape caustic in the central region near the location of the primary lens.  
We note that this caustic is nested and self-intersecting, and thus it appears to be different 
from the quadrilateral caustic of the binary 2L1S solution.  The source approached close to 
one of the cusps of this central caustic, producing the peak that occurred at $t_1$.  In the 
left panels of the figure, we mark three positions of the source, marked by $t_1$, $t_2$, 
and $t_3$, corresponding to the times of the three peaks in the light curve marked in 
Figure~\ref{fig:one}.

\subsection{Higher-order Effects}\label{sec:three-three}

We check the higher-order effects in the lensing light curve.  Considering these effects 
is important not only for precisely describing the light curve but also for constraining 
the physical lens parameters because the mass and distance to the lens are related to the 
microlens parallax.  In the modeling, we simultaneously consider both the microlens-parallax 
and lens-orbital effects because these effects can result in qualitatively similar deviations 
in lensing light curves \citep{Batista2011, Skowron2011, Han2016}.  To consider the 
lens-orbital motion of the close solution, we use the approximation that the $M_1$--$M_3$ 
binary pair is orbiting around their center of mass and the planetary companion $M_2$ is 
orbiting around $M_1$.  For the wide solution for which the binary companion, $M_3$, is 
located at a considerable distance from the primary, $M_1$, we consider only the orbital 
motion of the planetary companion around the primary lens component, $M_1$.

\begin{figure}
\includegraphics[width=\columnwidth]{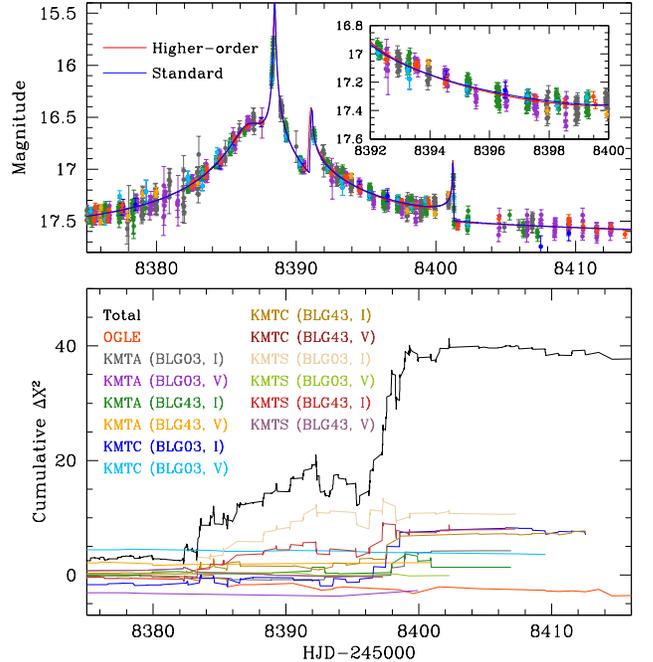}
\caption{
Comparison of models with and without the consideration of higher-order effects.  The 
model curves of the two solutions are presented in the upper panel, in which the red and 
blue curves are for the solutions with and without higher-order effects, respectively.  The 
lower panel shows the cumulative distributions of $\Delta\chi^2$ between the two solutions.  
The inset in the upper panel shows the zoomed region of $8392 \lesssim {\rm HJD}^\prime 
\equiv {\rm HJD}-2450000 \lesssim 8400$, during which a major fit improvement occurs.
The presented model light curves are  
for the solutions with $s_3<1.0$ and $u_0>0.0$.
\smallskip
}
\label{fig:five}
\end{figure}

In the lower panel of Figure~\ref{fig:five}, we present the cumulative distributions 
of $\chi^2$ difference between the two models obtained with and without considering 
the higher-order effects.  The black curve is for the total data.  The other curves 
are for the individual data sets, and the colors of the individual curves match those 
of the labels in the legend.  We note that the data taken from each KMTNet telescope 
are composed of two sets because the source is located in the two overlapping fields 
(BLG03 and BLG43 fields) that are directed with a slight offset to fill the gaps between 
the chips of the camera.  We also note that the MOA data set is not used for the 
higher-order modeling because of its relatively large photometric uncertainties.  The 
presented model is for the close solution with $s_3<1.0$ and $u_0>0.0$.  It is found 
that the consideration of the higher-order effects improves the fit by $\Delta\chi^2\sim 38$.  
We note that the other degenerate solutions result in similar fit improvement.  In the 
upper panel, we also present the model light curves obtained with (red curve) and without 
(blue curve) considering the higher-order effects.  In the inset of the upper-panel, 
we present zoomed view of the region of  $8392 \lesssim {\rm HJD}^\prime \lesssim 8400$, 
during which a major fit improvement occurs.

\begin{figure}
\includegraphics[width=\columnwidth]{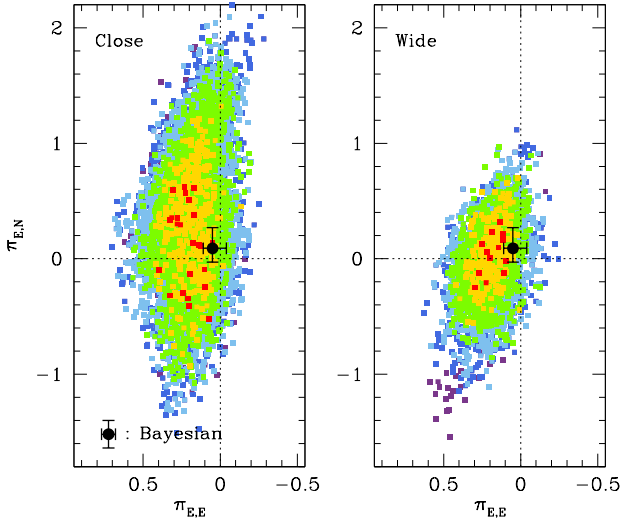}
\caption{
Distribution of $\Delta\chi^2$ of MCMC points in $\piee$--$\pien$ plane for the 
close (with $u_0>0.0$, left panel) and wide ($u_0>0.0$, right panel) solutions.  
The color coding is set to represent points within $1\sigma$ (red), $2\sigma$ (yellow), 
$3\sigma$ (green), $4\sigma$ (cyan), $5\sigma$ (blue), and $6\sigma$ (purple).  
In each 
panel, the point with error bars represents the ranges of the microlens-parallax parameters 
estimated from the Bayesian analysis.
\smallskip
}
\label{fig:six}
\end{figure}

We find that it is difficult to securely measure the higher-order effects.  The main 
reason for the difficulty is caused by the subtlety of the deviation induced by the 
effects.  This can be seen from the comparison of models with and without the effects 
presented in the upper panel of Figure~\ref{fig:five}, which shows that the two models 
result in very similar light curves.  Due to the subtle deviation, the uncertainties 
of the measured higher-order lensing parameters are very large.  In Figure~\ref{fig:six}, 
we present the $\Delta\chi^2$ distributions of MCMC points in the $\piee$--$\pien$ plane 
for the close (with $u_0>0.0$, left panel) and wide ($u_0>0.0$, right panel) solutions.  
The measured microlens-parallax parameters and their uncertainties are 
$(\pien, \piee)=( 0.18\pm 0.54, 0.23\pm 0.14)$, $(\pien, \piee)=(-0.12\pm 0.34, 0.11\pm 0.12)$ 
for the close and wide solutions, respectively. As we will discuss in Section~\ref{sec:five}, 
these error bars are far larger than the constraints of the Bayesian analysis.

\subsection{2L2S Analysis}\label{sec:three-four}

We additionally check solutions in which both the lens and source are binaries (2L2S).  In 
this modeling, we hold the trajectory of one source as that of the planetary 2L1S solution, 
which explains the peaks at $t_2$ and $t_3$, and test various trajectories of the other 
source to explain the peak at $t_1$. We find that the 2L2S modeling does not yield a solution 
that can explain the other peak at $t_1$, indicating that 2L2S model cannot explain all the 
anomalous features in the observed lensing light curve.

\section{Source Star}\label{sec:four}

We characterize the source star by estimating its de-reddened color, $(V-I)_0$, and 
brightness, $I_0$.  The de-reddened color and brightness are estimated from the instrumental 
values using the centroid of the red giant clump (RGC), for which its de-reddened color, 
$(V-I)_{\rm RGC, 0}$, and brightness, $I_{\rm RGC,0}$, are known, in the color-magnitude 
diagram (CMD) as a reference \citep{Yoo2004}.

\begin{figure}
\includegraphics[width=\columnwidth]{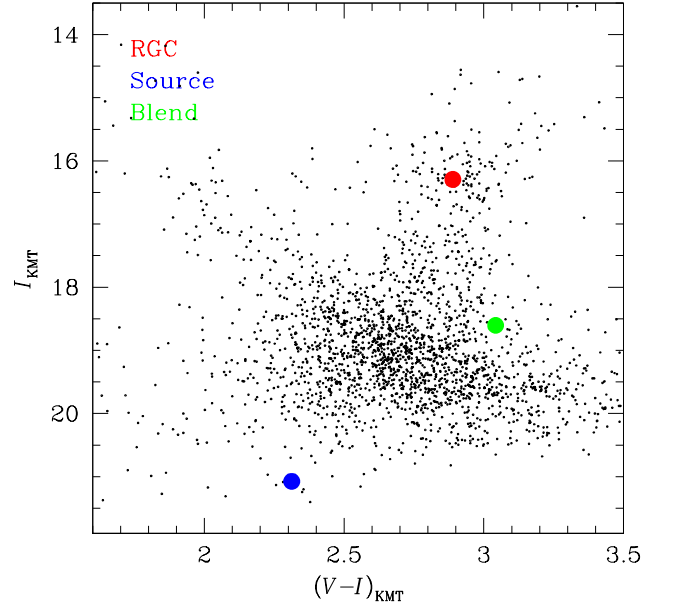}
\caption{
Positions of the source and blend with respect to the centroid of red giant clump (RGC) 
in the instrumental color-magnitude diagram constructed based on the pyDIA photometry 
of the KMTNet BLG03 $I$- and $V$-band data.
\bigskip
}
\label{fig:seven}
\end{figure}

In Figure~\ref{fig:seven}, we mark the position of the source in the instrumental CMD 
constructed based on the pyDIA photometry of the KMTC $I$- and $V$-band data. The 
instrumental color and brightness of the source are $(V-I, I) =(2.31\pm 0.03, 21.08\pm 0.01)$ 
compared to the RGC centroid values of $(V-I, I)_{\rm RGC}=(2.89, 16.29)$.  From the offsets 
in color and brightness between the source and RGC centroid together with the known de-reddened 
values $(V-I, I)_{\rm RGC,0}=(1.06, 14.35)$ of the RGC centroid \citep{Bensby2013,Nataf2013}, 
the de-reddened color and 
brightness of the source are estimated as $(V-I, I)_0=(0.47\pm 0.03, 18.96\pm 0.01)$.
The color and brightness indicate that the source is an F-type main-sequence star.

We determine the angular Einstein radius $\thetae$ and the relative lens-source proper 
motion $\mu$ from the angular source radius $\theta_*$ that is estimated from the 
measured source color.  For this, we first convert the measured $V-I$ color of the 
source into $V-K$ color using the color-color relation of \citet{Bessell1988}, and 
then estimate $\theta_*$ using the \citet{Kervella2004} relation between $V-K$ 
and $\theta_*$.  The estimated angular source radius is 
\begin{equation}
\theta_*= 0.37 \pm  0.03~\mu{\rm as}.
\label{eq1}
\end{equation}
With the measured value of $\theta_*$, the angular Einstein radius and the relative 
lens-source proper motions are estimated by 
\begin{equation}
\thetae= {\theta_* \over \rho} = 0.37 \pm  0.04~{\rm mas}.
\label{eq2}
\end{equation}
and 
\begin{equation}
\mu = {\thetae\over t_{\rm E}} = 3.13 \pm  0.30~{\rm mas}~{\rm yr}^{-1},
\label{eq3}
\end{equation}
respectively.

We find that the source star is unlikely to be located in the bulge and, instead, it is 
most likely to be located in the far disk behind the bulge.  According to the re-reddened 
color, $(V-I)_0\sim 0.47$, the source is an F-type star, but there are essentially no 
such bluish stars in the bulge. This indicates that the source is unlikely to be in the 
bulge and it should be located in the disk. A mid to late F-type star would be $\sim 3$--4 
magnitudes fainter than the clump giant if the source were located at the same distance 
as the clump giant. Considering that the source is $\sim 4.8$ magnitude fainter than the 
clump, it is likely that the source is located in the far disk behind the bulge.  The 
Galactic latitude of the source is $b=-2^\circ\hskip-2pt.47$.  Hence, the line of sight 
passes about 415~pc below the disk plane at a source distance of $D_{\rm S}\sim 10$~kpc. 
Considering that the disk scale height is $\sim 300$~pc, there would be some disk stars 
at this height, although the density is reduced.

Also marked in Figure~\ref{fig:seven} is the location of the blend.  The blend is $\sim 2.5$ 
magnitude brighter than the source.  We check the possibility of the lens being the blend 
itself as in the case of OGLE-2018-BLG-0740 \citep{Han2019b}.  For this, we measure the astrometric 
offset between the position of the baseline object, measured in the image obtained by 
combining 72 KMTC images taken before lensing magnification, and the position of the source, 
measured in the difference image obtained by combining 47 difference images taken during the 
lensing magnification.  The measured offset is $0.60$ pixels in the chip of the KMTNet camera, 
which corresponds to 0.22~arcsec.  This offset is much bigger than the astrometric errors in 
either the position of the ``baseline object'' (0.04 pixels) or the ``difference image'' 
(0.03 pixels).  Therefore, the blend must be due at least in part to an unrelated star or stars.

\section{Lens System}\label{sec:five}

For the unique determinations of the mass $M$ and distance $D_{\rm L}$ to the lens, it is 
required to measure both the microlens parallax $\pi_{\rm E}$ and the angular Einstein radius 
$\thetae$, which are related to $M$ and $D_{\rm L}$ by
\begin{equation}
M={\thetae\over \kappa\pi_{\rm E}};\qquad
D_{\rm L}={{\rm au}\over \pi_{\rm E}\thetae+\pi_{\rm S}},
\label{eq4}
\end{equation}
where $\kappa=4G/(c^2{\rm au})$ and $\pi_{\rm S}={\rm au}/D_{\rm S}$ is the parallax to the
source, and $D_{\rm S}$ denotes the distance to the source.  In the case of OGLE-2018-BLG-1700, 
the angular Einstein radius is measured, but the microlens parallax is not securely measured.
We, therefore, estimate the physical lens parameters by conducting a Bayesian analysis of the 
event based on the constraints of the measured event time scale and angular Einstein radius 
together with the constraint of the source location, i.e., far disk behind the bulge.

We conduct the Bayesian analysis using the prior models of the lens mass function and the physical 
and dynamical distributions of stars in the Galaxy.  Based on these models, we produce numerous 
artificial lensing events by conducting Monte Carlo simulation and construct the probability 
distributions of the lens mass and distance. In the analysis, we use the \citet{Chabrier2003} 
model for the mass function of stars and the \citet{Gould2000} model for the mass function of 
stellar remnants.  For the physical and dynamical distributions of matter, we use the 
\citet{Han2003} and \citet{Han1995} models, respectively. 
Among the produced events, the probability distributions are constructed
for events with time scales and angular Einstein radii lying within the uncertainty ranges of
the measured $t_{\rm E}$ and $\thetae$, with disk source stars lying at distances
$D_{\rm S} \geq 8$ kpc.
From the constructed probability distributions, we then choose the physical parameters
as the median values and the uncertainties are estimated as the 68\% ranges of the distributions.

In the Monte Carlo simulation, we model the lens distribution as that of the bulge.
Because the source lies in the far disk, the lens could in principle lie in either the far
disk, the bulge or the near disk. However, the observed proper motion of
$\mu_{\rm rel}=3.1\pm 0.3~{\rm mas~yr}^{-1}$ virtually rules out near-disk lenses for which
the expected mean proper motion would be
$\langle\muvec_{\rm rel}\rangle \simeq 2({\bf v}_{\rm rot}/v_{\rm rot}^2)/D_{\rm S}
\rightarrow 9.3~{\rm mas~yr}^{-1}$
in the direction of Galactic rotation.
Only improbably large peculiar motions of the lens or source (relative to the mean
circular motion of the Galactic disk) could then bring $\mu_{\rm rel}$ within the observed
range. While far-disk lenses could in principle satisfy the proper-motion constraint, the
physical-matter distribution along the line of sight (and beyond the near disk) is
completely dominated by the bulge. We therefore model the lens distribution
as that of the bulge.

\begin{figure}
\includegraphics[width=\columnwidth]{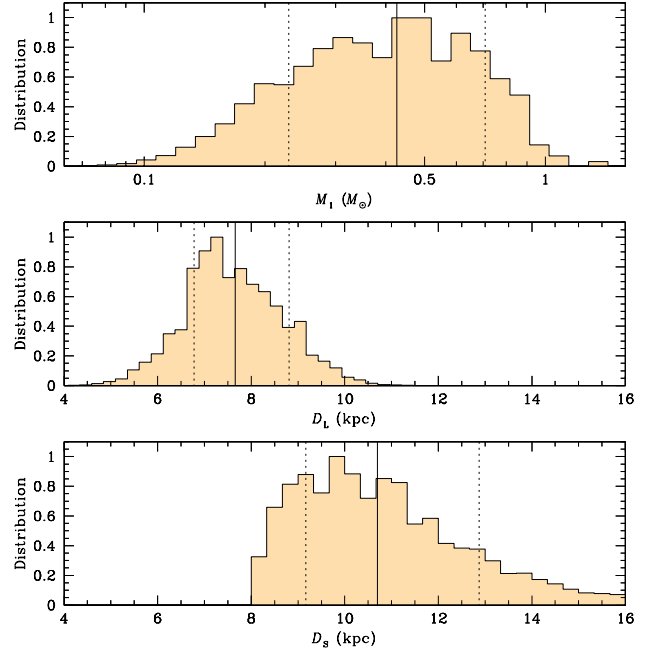}
\caption{
Probability distributions of the primary-lens mass $M_1$ (top panel), distance to the lens 
$D_{\rm L}$ (middle panel), and the distance to the source $D_{\rm S}$ (bottom panel) 
obtained from the Bayesian analysis.  For each distribution, the solid vertical line 
represents the median value and the two dotted lines represent the 68\% range of the 
distribution.  The distributions are for the wide solution and the close solution results 
in nearly identical distributions.
\smallskip
}
\label{fig:eight}
\end{figure}

In Figure~\ref{fig:eight}, we present the probability distributions of the primary-lens mass, 
$M_1$, distance to the lens, $D_{\rm L}$, and distance to the source, $D_{\rm S}$, obtained 
from the Bayesian analysis.  In Table~\ref{table:two}, we summarize the physical lens parameters,
including the masses of the individual lens components ($M_1$, $M_2$, and $M_3$), distances
to the lens and source ($D_{\rm L}$ and $D_{\rm S}$), and the projected physical separations
of $M_2$ and $M_3$ measured from the position of $M_1$
($a_{\perp,1-2}$ and $a_{\perp,1-2}$).

We find that the result from the Bayesian analysis is consistent with the 
microlens-parallax measurement. 
For the comparison of the parallax distributions, we compute the north and east 
components of the microlens parallax vectors $\pivec_{\rm E}$ 
of events produced by the Bayesian analysis as
\begin{equation}
\begin{array}{lr}
\pien = &  \pi_{{\rm E},b} \cos\gamma + \pi_{{\rm E},l} \sin\gamma,  \\
\piee = & -\pi_{{\rm E},b} \sin\gamma + \pi_{{\rm E},l} \cos\gamma,
\label{eq5}
\end{array}
\end{equation}
respectively. 
Here $\gamma=60.3^\circ$ represents the angle between arcs of constant Galactic latitude $(b)$ 
and constant equatorial declination $(\delta)$.
The microlens-parallax components along the galactic longitude ($l$) 
and latitude ($b$) directions are computed from the relative lens-source transverse velocity 
vector ${\bf v}=(v_l, v_b)$ by
\begin{equation}
\begin{array}{lr}
\pi_{{\rm E},l}= & \pie (v_l/v) ,  \\
\pi_{{\rm E},b}= & \pie (v_b/v) .  \\
\end{array}
\label{eq6}
\end{equation}
In Figure~\ref{fig:six}, we mark the ranges of $\pien$ and $\piee$ estimated from the 
Bayesian analysis as a dot with error bars superposed on the $\Delta\chi^2$ distribution 
of MCMC points obtained from light curve fitting.
It is found that the Bayesian result is consistent with the microlens parallax 
measurement, although the measurement uncertainty of $\pie$ is large.

\begin{deluxetable}{lcc}
\tablecaption{Best-fit lensing parameters\label{table:two}}
\tablewidth{240pt}
\tablehead{
\multicolumn{1}{c}{Parameter}            &
\multicolumn{1}{c}{Wide ($s_3>1.0$)}     &
\multicolumn{1}{c}{Close ($s_3<1.0$)}    \\
\multicolumn{1}{c}{}                     &
\multicolumn{1}{c}{Circumstellar}     &
\multicolumn{1}{c}{Circumbinary}
}
\startdata                                              
$M_1$ ($M_\odot$)        &  $0.42^{+0.29}_{-0.19}$  &  $\leftarrow$ \\
$M_2$ ($M_{\rm J}$)      &  $4.40^{+3.04}_{-2.00}$  &  $\leftarrow$ \\
$M_3$ ($M_\odot$)        &  $0.12^{+0.08}_{-0.05}$  &  $\leftarrow$\\
$D_{\rm L}$ (kpc)        &  $7.6^{+1.2}_{-0.9}$     &  $\leftarrow$ \\
$D_{\rm S}$ (kpc)        &  $10.7^{+2.2}_{-1.5}$    &  $\leftarrow$ \\
$a_{\perp,1-2}$ (au)     &  $2.8^{+3.2}_{-2.5}$     &  $\leftarrow$\\
$a_{\perp,1-3}$ (au)     &  $10.5^{+12.1}_{-9.2}$   &  $0.75^{+0.87}_{-0.66}$ 
\enddata                            
\tablecomments{
$M_1$, $M_2$, and $M_3$ represent the masses of the individual triple-lens components, 
$D_{\rm L}$ and $D_{\rm S}$ denote the distances to the lens and source, respectively, 
and $a_{\perp,1-2}$ and $a_{\perp,1-3}$ represent the projected physical separations of
between $M_1$--$M_2$ and $M_1$--$M_3$ pairs, respectively.  The ``$\leftarrow$'' symbols 
for the close solution imply that the values are the same as for the wide solution.
\smallskip
}
\end{deluxetable}

The interpretation of the planetary orbit varies depending on the solutions.  According to 
the ``wide solution'' with $s_3>1.0$, the planet has an S-type orbit, in which the planet orbits 
around one of the two stellar binary stars, i.e., circumstellar planet. According to the ``close 
solution'', on the other hand, the planet has a P-type orbit, in which the planet orbits around 
the barycenter of the close binary stars, i.e., circumbinary planet. 
The planet is a super-Jupiter with a mass of 
\begin{equation}
M_2=4.40^{+3.04}_{-2.00}~M_{\rm J},
\label{eq7}
\end{equation}
and the stellar binary components are early and late M-type dwarfs with masses
\begin{equation}
M_1=0.42^{+0.29}_{-0.19}~M_\odot
\label{eq8}
\end{equation}
and 
\begin{equation}
M_3=0.12^{+0.08}_{-0.05}~M_\odot, 
\label{eq9}
\end{equation}
respectively. 
The projected $M_1$--$M_2$ separation is 
\begin{equation}
a_{\perp,1-2} = 
2.8^{+3.2}_{-2.5}~{\rm au}
\label{eq10}
\end{equation}
for both the close and wide solutions. However, the projected $M_1$--$M_3$ separation 
estimated from the close solution, 
\begin{equation}
a_{\perp,1-3} = 0.75^{+0.87}_{-0.66}~{\rm au}\qquad ({\rm close}), 
\label{eq11}
\end{equation}
is greatly different from the separation of
\begin{equation}
a_{\perp,1-3}= 10.5^{+12.1}_{-9.2}~{\rm au} \qquad ({\rm wide})
\label{eq12}
\end{equation}
estimated from the wide solution.  The distance to the lens is 
\begin{equation}
D_{\rm L}=7.6^{+1.2}_{-0.9}~{\rm kpc},
\label{eq13}
\end{equation}
and the source is estimated to be in the far disk at a distance of 
\begin{equation}
D_{\rm S}=10.7^{+2.2}_{-1.5}~{\rm kpc}.
\label{eq14}
\end{equation}

\section{Summary and Conclusion}\label{sec:six}

We found a planet belonging to a stellar binary system from the analysis of the 
microlensing event OGLE-2018-BLG-1700.  We identified the triple nature of the 
lens from the fact that the complex anomaly pattern could be decomposed into two 
parts produced by two binary-lens events, in which one binary pair had a very low 
mass ratio between the lens components and the other pair had similar masses.  We 
found two sets of degenerate solutions, in which one solution had a projected separation 
between the stellar lens components less than the angular Einstein radius $\thetae$, while 
the other solution had a separation greater than $\thetae$.  In order to estimate the 
physical lens parameters, we conducted a Bayesian analysis with the constraints of the 
measured event time scale and angular Einstein radius together with the location of the 
source lying in the far disk behind the bulge.  From this, we found that the planet was 
a super-Jupiter with a mass of $4.4^{+3.0}_{-2.0}~M_{\rm J}$, and the stellar binary 
components were early and late M-type dwarfs with masses $0.42^{+0.29}_{-0.19}~M_\odot$ 
and $0.12^{+0.08}_{-0.05}~M_\odot$, respectively.  The interpretation of the planetary 
orbit varied depending on the solutions and the planet was a circumstellar planet orbiting 
around one of the two binary stars according to the wide solution, while it was a 
circumbinary planet orbiting around the center of mass of the binary stars according to 
the close solution.

\acknowledgments
Work by CH was supported by the grant (2017R1A4A1015178) of National Research Foundation of Korea.
Work by AG was supported by US NSF grant AST-1516842 and by JPL grant 1500811.
AG received support from the European Research
Council under the European Union's Seventh Framework
Programme (FP 7) ERC Grant Agreement n.~[32103].
The OGLE project has received funding from the National Science Centre, Poland, grant
MAESTRO 2014/14/A/ST9/00121 to AU.
This research has made use of the KMTNet system operated by the Korea
Astronomy and Space Science Institute (KASI) and the data were obtained at
three host sites of CTIO in Chile, SAAO in South Africa, and SSO in
Australia.
The MOA project is supported by JSPS KAKENHI Grant Number JSPS24253004,
JSPS26247023, JSPS23340064, JSPS15H00781, JP17H02871, and JP16H06287.
YM acknowledges the support by the grant JP14002006.
DPB, AB, and CR were supported by NASA through grant NASA-80NSSC18K0274. 
The work by CR was supported by an appointment to the NASA Postdoctoral Program at the Goddard 
Space Flight Center, administered by USRA through a contract with NASA. NJR is a Royal Society 
of New Zealand Rutherford Discovery Fellow.
We acknowledge the high-speed internet service (KREONET)
provided by Korea Institute of Science and Technology Information (KISTI).

\end{document}